\documentstyle[preprint,revtex]{aps}
\begin{document}
\draft
\preprint{SNUTP-93/82}
\preprint{Revised}
\begin{title}
Non-Abelian Chern-Simons Quantum Mechanics
\end{title}
\begin{title}
and
\end{title}
\begin{title}
Non-Abelian
Aharonov-Bohm Effect \end{title}
\author{Taejin Lee\cite{tlee}}
\begin{instit}
Department of Physics, Kangwon National University, Chuncheon 200-701, KOREA
\end{instit}
\author{Phillial Oh\cite{poh}}
\begin{instit}
Department of Physics, Sung Kyun Kwan University, Suwon 440-746, KOREA
\end{instit}
\begin{abstract}

We construct a classical action for a system of $N$ point-like
sources which carry SU(2) non-Abelian charges coupled to  non-Abelian
Chern-Simons gauge fields, and develop a quantum mechanics for them.
Adopting the coherent state quantization and solving the Gauss'
constraint in an
appropriately chosen gauge, we obtain a quantum mechanical
Hamiltonian given in
terms of the  Knizhnik-Zamolodchikov connection. Then we study the
non-Abelian Aharonov-Bohm effect, employing the obtained Hamiltonian
for two-particle sector. An explicit evaluation of the differential
cross section for the  non-Abelian Aharonov-Bohm scattering is given.

\end{abstract}


\narrowtext

\section{Introduction}
\label{sec:intro}

Since Aharonov and Bohm discussed in their celebrated paper \cite{ahbo}
the significance of the phase that arises from the
charge-flux interaction, now known as the Aharonov-Bohm effect, it has been
one of the most important subjects in both experimental and theoretical
physics \cite{peshkin}. The subject spreaded diverse branches and recently
has developed into the anyon physics \cite{anyon} which has wide range
of applications, including the fractional quantum Hall
effect \cite{frac} and the high-$T_c$ super-conductivity \cite{high}.

In this paper, we discuss in some detail the latest development in this
direction, that is, the non-Abelian Chern-Simons quantum mechanics
\cite{lo1,lo2} and the non-Abelian Aharonov-Bohm effect
\cite{wilwu,ver}. Recently we proposed a classical action for a system
of the non-Abelian Chern-Simons (NACS) particles and developed a
quantum mechanical description for them. The NACS particles are
point-like sources which carry non-Abelian charges and interact with
each other through the non-Abelian Chern-Simons term \cite{des}. Their
interaction is essentially the non-Abelian Aharonov-Bohm effect, just
as the interaction between anyons is the Aharonov-Bohm effect.

The non-Abelian Aharonov-Bohm effect, however, is not a newly
raised issue: It was considered long ago by Wu and Yang \cite{wuyang} and its
details were studied in a test-particle framework by Horv\'athy \cite{hor}.
Wu and Yang proposed a gedanken experiment to test the existence of the
$SU(2)$ isotopic spin gauge field: They consider a scattering of the
beam of protons, neutrons or their mixture around a cylinder where an
isotopic magnetic flux is confined.  A rotating cylinder made of heavy
elements with a neutron excess was suggested as a source of the
non-Abelian magnetic flux. The existence of the isotopic $SU(2)$ gauge
field would be manifested through the non-integrable phase factor, upon
which the outcome of experiment depends.

Recently it has been in the limelight again as it becomes clear that the
NACS particle and the non-Abelian Aharonov-Bohm effect might be realized in
various physical phenomena such as the fractional quantum Hall effect,
cosmic strings \cite{cosmic}, and the gravitational scattering in (2+1)
dimensions \cite{3dgr}.

The NACS particles are in a sense non-Abelian generalization of anyons. Their
physical properties are similar to those of anyons, but can be certainly
distinguished from. While anyons acquire fractional spins and satisfy
exotic statistics, the NACS particles acquire fractional but rational
spins and exhibit yet more generalized braid (or non-Abelian) statistics
\cite{froh}. In this respect, it is quite interesting to apply the NACS
particles to those physical phenomena, in some generalized context, where
anyons play important roles, and to scrutinize the outcome.
Such attempts have
been made in refs.\cite{wen,moo,balfra} where the possibility for the NACS
particles to be realized in the fractional quantum Hall effect, the
paradigm of anyons, is discussed.

The NACS particles may manifest their existence also in somewhat different
circumstances. Such examples include the vortices in (2+1) dimensions
and the cosmic strings in (3+1) dimensions which are formed when some
gauge group $G$ is broken down via Higgs mechanism to some discrete
non-Abelian subgroup $H$ \cite{discrete}. Since interactions between
these objects are characterized by the holonomies, equivalents of Wu
and Yang's non-integrable phase factors, associated with the windings
around themselves, their interactions are effectively non-Abelian
Aharonov-Bohm effect.     The point-like sources which
gravitationally interact with each other in (2+1) dimensions also
belong to the examples. Since the (2+1) dimensional gravity is
equivalent to the non-Abelian Chern-Simons theory of group $ISO(2,1)$
\cite{wittgr}, they are NACS particles with the internal symmetry group
of $ISO(2,1)$. Accordingly the (2+1) dimensional gravity can be also
discribed in the framework of the non-Abelian Aharonov-Bohm effect
\cite{3dgr}.

The rest of the paper is organized as follows. In Sec. II we discuss how to
construct a classical action describing non-Abelian
Chern-Simons particles and
analyze its constraint algebra. In Sec. III we apply the coherent state
quantization to the Chern-Simons gauge fields and introduce a gauge fixing
condition, called holomorphic gauge, where the Gauss' constraints are
explicitly solved. Then in Sec. IV we obtain a quantum mechanical
description
of the NACS particles and show that they obey braid statistics. Sec. V is
devoted to the two body problem in non-Abelian Chern-Simons
quantum mechanics.
In Sec. VI we evaluate the differential cross section of the
non-Abelian Aharonov-Bohm scattering. In Sec VII we conclude the paper with a
brief summary and a discussion on applications of
the non-Abelian Chern-Simons quantum mechanics.

\section{Non-Abelian Chern-Simons Particles}
\label{sec:nacsp}

A classical description of point-like sources with non-Abelian (isospin)
charges can be given in terms of their spatial coordinates and the isospin
vectors which transform under the adjoint representation of the internal
symmetry group. Endowing them with the corresponding non-Abelian magnetic
fluxes to make them the NACS particles is done by introducing the
non-Abelian Chern-Simons term and minimally coupling their isospin
charges with the Chern-Simons gauge fields.
To be specific and to avoid unnecessary burden, we take
the internal symmetry group to be $SU(2)$ throughout the paper. (For
discussions on the NACS particles with $SU(N)$, $N\ge 2$ isospin charges,
consult ref. \cite{lo2}.)

Although there are several equivalent ways \cite{bal78,bal90,alek}
to define the
isospin degrees of freedom, we find it convenient to
define the isospin vectors
directly on the reduced phase space $S^2$ for the $SU(2)$ internal symmetry
group as follows \cite{oh}
\begin{equation}
Q^1_\alpha= J_\alpha \sin \theta_\alpha \cos\phi_\alpha,\quad Q^2_\alpha=
J_\alpha\sin \theta_\alpha \sin \phi_\alpha,\quad Q^3_\alpha =
J_\alpha\cos\theta_\alpha \label{iso}
\end{equation}
where $\theta_\alpha,\, \phi_\alpha$ are the coordinates
of the internal $S^2$ and $J_\alpha$ is a constant.
Denoting the spatial coordinates of the particles by
${\bf q}_\alpha$, $\alpha=1,2,\cdots,N$, we can construct the classical
Lagrangean as
\[ L = \sum_\alpha\left(-{1 \over 2} m_\alpha \dot{\bf
q}_\alpha^2 +J_\alpha \cos \theta_\alpha \dot{\phi}_\alpha\right)
-\kappa\int
d^2 {\bf x} \,\epsilon^{\mu\nu\lambda} {\rm tr}\left(A_\mu \partial_\nu
A_\lambda +{2\over 3} A_\mu A_\nu A_\lambda\right) \]
\begin{equation}
+\int d^2{\bf x}\sum_\alpha \left(A^a_i(t,{\bf x}) \dot
q^i_\alpha +A^a_0(t, {\bf x})\right) Q^a_\alpha \delta ({\bf
x}-{\bf q}_\alpha)\label{lag}
\end{equation}
Here $4\pi\kappa = {\rm integer}$, $A_\mu=A_\mu^a T^a$,
$[T^a, T^b] = \epsilon^{abc} T^c$, ${\rm tr} (T^a T^b) = -1/2 \delta^{ab}$,
and the space-time signature is $(+,-,-)$.

With the defining equations
for the isovectors Eq.(\ref{iso}), we obtain the following Euler-Lagrangean
equations
\begin{equation}
m_\alpha \ddot{q}_{\alpha i} = - (F^a_{ij}({\bf q}_\alpha)
\dot{q}_\alpha^j + F^a_{i0}({\bf q}_\alpha)) Q^a_\alpha, \label{eul1}
\end{equation}
\begin{equation}
\dot{Q}^a_\alpha = -\epsilon^{abc} (A^b_i({\bf q}_\alpha) \dot{q}^i_\alpha+
A^b_0({\bf q}_\alpha)) Q^c_\alpha, \label{eul2}
\end{equation}
\begin{equation}
{\kappa \over 2} \epsilon^{ij} F^a_{ij} ({\bf x}) = -\sum_\alpha Q^a_\alpha
\delta({\bf x}-{\bf q}_\alpha), \label{gau}
\end{equation}
\begin{equation}
\kappa \epsilon^{ij} F^a_{j0}({\bf x})
= -\sum_\alpha Q^a_\alpha \dot{q}^i_\alpha \delta({\bf x}-{\bf
q}_\alpha) \label{eul4}
\end{equation}
where $F^a_{ij}= \partial_i A^a_j -\partial_j A^a_i
+\epsilon^{abc} A^b_i A^c_j$. The first two equations are Wong's equations
\cite{wong} and the third one corresponds to the Gauss' law constraint which
tells us that the NACS particle of isospin charge $Q^a_\alpha$ carries the
magnetic flux $-Q^a_\alpha/\kappa$;
\begin{equation}
\Phi_{\rm m} = \frac{1}{2}\int_{B_\alpha}
\epsilon^{ij}F^a_{ij}({\bf x}) d^2x = -\frac{1}{\kappa} Q^a_\alpha
\end{equation}
where $B_\alpha$ denotes a small patch covering the position of the
$\alpha$-th NACS
particle.

Since the part of Lagrangean for the isospin degrees of freedom is usually
of first order, second class constraints arise in the procedure of
canonical quantization. These second class constraints are entangled
with the first class constraints which generate gauge symmetries and
it makes the constraint analysis difficult \cite{bal78,bal90}. One of the
advantages we have, to define the isovectors as in Eq.(\ref{iso}) is
that it is easy to avoid the second class constraints from
the outset by judiciously defining the Poisson bracket \cite{ber,fadd}.
Introducing the canonical momenta $p^i_\alpha$,
\begin{equation}
p^i_\alpha ={\partial L \over \partial
\dot q_{i\alpha}}= m_\alpha \dot{q}^i_\alpha+ A^{a i} ({\bf q}_\alpha)
Q^a_\alpha
\end{equation}
we can
convert the given Lagrangean Eq.(\ref{lag}) to a first order Lagrangean
\[ L = \sum_\alpha\left(p^i_\alpha \dot q_{i\alpha} + J_\alpha
\cos \theta_\alpha \dot \phi_\alpha\right) +\int d^2{\bf x}\left(
{\kappa\over 2}\epsilon^{ij}\dot A^a_i A^a_j\right) - H, \]
\[ H  = H_0 -\int d^2 {\bf x} \Biggl[  A^a_0\left( {\kappa \over 2}
\epsilon^{ij} F^a_{ij} +\sum_\alpha Q^a_\alpha \delta({\bf x}- {\bf
q}_\alpha)\right)\Biggr]  \]
\begin{equation}
H_0 = \sum_\alpha {1\over 2 m_\alpha}\left(p^i_\alpha-A^{ai}({\bf q}_\alpha)
Q^a_\alpha\right)^2. \label{first}
\end{equation}

This first order Lagrangean can be rewritten as
\cite{fadd}
\begin{equation}
L= a_I({\xi})\dot\xi^I -H(\xi)\label{forder}
\end{equation}
and the Euler-Lagrangean equations from Eq.(\ref{forder}) as
\[ f_{IJ}(\xi)\dot\xi^J = \frac{\partial}{\partial \xi^I} H(\xi),
\]
\[ f_{IJ}(\xi)=\frac{\partial}{\partial \xi^I}
a_J(\xi)-\frac{\partial}{\partial \xi^J}
a_I(\xi)\]
where $\xi^I$ denote collectively the canonical variables,
$p^i_\alpha$, $q^i_\alpha$, $\theta_\alpha$, $\phi_\alpha$, $A^a_i({\bf x})$.
$f_{IJ}(\xi)$ defines the pre-symplectic two form $f$ by
\begin{equation}
f=\frac{1}{2}f_{IJ}(\xi)d\xi^I d\xi^J = da(\xi)
\end{equation}
where $a(\xi)$ is the canonical one form, $a(\xi) = a_I(\xi)d\xi^I$.
For the given first order Lagrangean Eq.(\ref{first}), we find
\begin{equation}
f = dp^i dq^i - J\sin \theta d\theta d\phi -\int d^2{\bf x}\frac{\kappa}{2}
\epsilon^{ij} \delta A^a_i({\bf x}) \delta A^a_j ({\bf x}) \label{pre}
\end{equation}
where we suppress the indices labeling the particles.
When the matrix $(f_{IJ})$ is non-singular, i.e., has
its inverse $(f^{IJ})$ as
in Eq.(\ref{pre}), the Poisson bracket is taken as
\begin{equation}
\{F(\xi), G(\xi)\} = f^{IJ}(\xi)\frac{\partial F(\xi)}{\partial \xi^I}
\frac{\partial G(\xi)}{\partial \xi^J}\label{poiss}
\end{equation}
and the Euler-Lagrangean equations can be expressed as
\begin{equation}
\dot \xi^I = \{ \xi^I, H\} = f^{IJ} \frac{\partial}{\partial \xi^J} H.
\end{equation}

 From Eqs.(\ref{pre}) and (\ref{poiss}), we can define
the Poisson bracket as
\[ \{F, G\}= \sum_\alpha\left[\left({\partial F
\over \partial q^i_\alpha}{\partial G \over \partial p_{\alpha i}}-
{\partial F \over \partial p_{\alpha i}}{\partial G \over \partial
q^i_\alpha}\right) -{1\over J_\alpha\sin\theta_\alpha}\left({\partial F
\over \partial \phi_\alpha}{\partial G \over \partial \theta_\alpha}-
{\partial F \over \partial \theta_\alpha}{\partial G \over \partial
\phi_\alpha}\right)\right] \]
\begin{equation}
 +\int d^2 {\bf x} \left({1\over\kappa}\epsilon_{ij}
{\delta F \over \delta A^a_i
({\bf x})}{\delta G \over \delta A^a_j ({\bf x})}\right)\label{poi}
\end{equation}
and the fundamental commutators as
\[ \{q^i_\alpha, p_{\beta j}\} = \delta^i_j\delta_{\alpha\beta},\qquad
\{Q^a_\alpha,Q^b_\beta\} =\epsilon^{abc}
Q^c_\alpha\delta_{\alpha\beta}\]
\begin{equation}
\{A^a_i({\bf x}), A^b_j({\bf y})\} = {1\over \kappa}\epsilon_{ij}
\delta({\bf x}- {\bf y})\delta^{ab}.
\end{equation}

As expected,  the Gauss' law constraints
\begin{equation}
\Phi^a = {\kappa \over 2}\epsilon^{ij} F^a_{ij} ({\bf x}) +
\sum_\alpha Q^a_\alpha \delta({\bf x}- {\bf q}_\alpha) =
0\label{gaus}
\end{equation}
form $SU(2)$ algebra and no further constraints arise
\begin{equation}
\{\Phi^a ({\bf x}),\Phi^b ({\bf y})\} =
\epsilon^{abc} \Phi^c\delta({\bf x}-{\bf y}), \qquad
\{H_0, \Phi^a ({\bf x})\} = 0.
\end{equation}
In contrast to the approach of Balachandran {\it et al.} \cite{bal78,bal90},
the second class constraints never appear in ours.

If one can solve the Gauss' constraints explicitly, one can describe the
dynamics of the NACS particles solely by the quantum mechanical Hamiltonian
$H_0$ in Eq.(\ref{first}) with $A^a_i({\bf q}_\alpha)$ which
are determined through the Gauss' constraints. Thus it is desirable to
solve the Gauss' constraints explicitly, if possible.
We may attempt to solve the constraints in Coulomb or covariant gauges, but
find it difficult. However, as
we shall see, there are two gauge conditions where the constraints have
explicit solutions. The first one is the axial gauge condition. Considering
the nature of (2+1) dimensions, one can easily understand how the axial gauge
works. Choosing an axial gauge condition, say, $A^a_1=0$, we get a solution
for the Gauss' constraint \cite{gua}
\begin{equation}
A^a_1({\bf x})=0,\quad A^a_2({\bf x})=-\frac{1}{\kappa}\sum_\alpha
Q^a_\alpha\theta(x-x_\alpha)\delta(y-y_\alpha)+f(y) \label{axial}
\end{equation}
where $f(y)$ is an arbitrary function of $y$. Unfortunately, it is rather
awkward to describe the dynamics of the NACS particles with the axial gauge
solution because of the strings attached on them.
(Analyzing the the well-known anyon quantum mechanics in the axial
gauge  \cite{kapu} may give us an idea. But we will not pursue it here.)
The alternative is the holomorphic gauge condition which we shall discuss
in the following section.

\section{Coherent State Quantization and Holomorphic Gauge}
\label{sec: coherent}

If one adopts the coherent state quantization
\cite{klauder} for the gauge fields, one can choose a
better gauge condition. We recall that $A^a_1$ and $A^a_2$
are canonical conjugates to each other, i.e.,
$\{A^a_1({\bf x}), A^b_2({\bf y})\}
=\kappa\delta^{ab}\delta({\bf x}-{\bf y})$, or upon
quantization $[\hat A^a_1({\bf x}), \hat A^b_2({\bf y})]
=i\kappa\delta^{ab}\delta({\bf x}-{\bf y})$. This suggests
for us to define `creation' and `annihilation' operators
\begin{equation} {\cal A}^{a \dagger} =
\sqrt{\frac{1}{2\kappa}}\left(\hat A^a_1 -i \hat
A^a_2\right),\quad {\cal A}^{a} =
\sqrt{\frac{1}{2\kappa}}\left(\hat A^a_1 +i \hat
A^a_2\right), \end{equation} \[ \left[{\cal A}^{a}({\bf
x}), {\cal A}^{b \dagger} ({\bf y})\right]
=\delta^{ab}\delta({\bf x}-{\bf y})  \] and to construct
coherent states \begin{equation}
\vert A_{\bar z}> \equiv \exp\left(\sqrt{\frac{\kappa}{2}}\int d^2 {\bf x}
A^a_{\bar z} {\cal A}^{a\dagger}\right)\vert 0>
\end{equation}
and their adjoints
\begin{equation}
<A_z\vert \equiv <0\vert \exp\left(\sqrt{\frac{\kappa}{2}}\int d^2 {\bf x}
A^a_{z} {\cal A}^{a}\right).
\end{equation}
Then we have
\begin{equation}
<A_z\vert A_{\bar z}> = \exp\left(\frac{\kappa}{2}\int d^2{\bf x}
A^a_z A^a_{\bar z}\right)
\end{equation}
and the resolution of identity
\begin{equation}
\int D A_z D A_{\bar z} \exp\left(-\frac{\kappa}{2}\int d^2{\bf x} A^a_z
A^a_{\bar z}\right)\vert A_{\bar z}><A_z\vert = I.\label{resol}
\end{equation}
Partitioning the time interval $[t_1, t_2]$ into many pieces and repeatedly
using the resolution of identity Eq.(\ref{resol}), we obtain a functional
integral representation for the transition amplitude (in the sector of
gauge fields)
\begin{equation}
<A_z^f, t_f\vert A_{\bar z}^i, t_i>=
\int D A_z D A_{\bar z} \exp\Biggl\{\frac{\kappa}{2}
\int d^2{\bf x} \left(A^f_z
A^f_{\bar z} +A^i_z A^i_{\bar z} \right)+\label{trgauge}
\end{equation}
\[
i\int d^2{\bf x} \int^{t_2}_{t_1} dt
\left[\frac{i\kappa}{2}\left(A^a_{z} \dot A^a_{\bar z}
-\dot A^a_{z} A^a_{\bar
z}\right) -H(A_z, A_{\bar z})\right]\Biggr\}
\]
where $H(A_z, A_{\bar z})$ is obtained by
substituting ${\cal A}^{a\dagger}\rightarrow A^a_z$, ${\cal A}^{a}\rightarrow
A^a_{\bar z}$ in the given Hamiltonian $H({\cal A}^{a\dagger}, {\cal
A}^{a})$. It is worthwhile to note that $A^a_z$ and $A^a_{\bar z}$ must be
treated as independent variables \cite{brown}.

In order to get the path integral
representation of the physical transition amplitude $Z$, now we include the
particle sector
\begin{equation}
Z=\int D p^z D q^{\bar z} D p^{\bar z} D
q^z D\cos\theta D\phi  D A_z D A_{\bar z} D A_0\label{ampli}
\end{equation}
\[
\exp\left\{
-\kappa i \int d^2 z\left(A^f_{\bar z} A^f_z+A^i_{\bar z} A^i_z\right)\right\}
\exp\{i\int^{t_f}_{t_i} dt L\},
\]
\[
L= \sum_\alpha\left(p^{\bar z}_\alpha \dot z_{\alpha}+
p^{z}_\alpha \dot{{\bar z}}_{\alpha}+ J_\alpha \cos\theta_\alpha
\dot\phi_\alpha\right) +\int d^2 z \left(
{\kappa\over 2}\left(\dot A^a_z A^a_{\bar z} -\dot A^a_{\bar z} A^a_{
z}\right) +  A^a_0 \Phi^a \right) - H, \]
\[
H = \sum_\alpha {2\over
m_\alpha}\left(p^{\bar z}_\alpha-A^{a}_z(z_\alpha, \bar z_\alpha)
Q^a_\alpha\right) \left(p^{z}_\alpha-A^{a}_{\bar z}(z_\alpha, \bar z_\alpha)
Q^a_\alpha\right)
\]
where $\Phi^a(z)=\kappa F^a_{z\bar z}+\sum_\alpha
Q^a_\alpha\delta (z-z_\alpha)=0.$ For the sake of convenience, we employed in
Eq.(\ref{ampli}) complex coordinates where
\[ z = x+ iy,\quad \bar z = x- iy, \]
\[ z_\alpha = q^1_\alpha + i q^2_\alpha,\quad \bar z_\alpha = q^1_\alpha - i
q^2_\alpha.\]
Adopting the coherent state quantization
for the Chern-Simons gauge fields \cite{ems,bos}
effectively extends the gauge
orbit space in that $A^a_z$ and $A^a_{\bar z}$ are treated as independent
variables \cite{brown}.
It implies concurrently that there are wider class of
gauge fixing conditions available in the frame-work of the coherent state
quantization: We can choose, $A^a_{\bar z}=0$, as a gauge fixing condition.
This gauge fixing condition may be called holomorphic gauge condition.
In this gauge,  the Gauss' constraint reduces to
\begin{equation}
\Phi^a(z)=-\kappa \partial_{\bar z} A^a_z +\sum_\alpha
Q^a_\alpha\delta(z-z_\alpha)=0 \label{gauss}
\end{equation}
and has an explicit solution
\begin{equation}
A^a_{\bar z} (z, \bar z)= 0,\quad
A^a_z (z, \bar z) = {i\over 2\pi \kappa}\sum_\alpha  Q^a_\alpha
{1\over z -z_\alpha}\label{sol}
\end{equation}
which is less singular that that in the axial gauge Eq.(\ref{axial}).
As we shall see, it is amusing to observe that the dynamics of the NACS
particles lies  entirely in the holomorphic sector of $z$ in the holomorphic
gauge.

Reaching this point, one may realize that the holomorphic gauge may not be
connected to the more conventional gauges such as Coulomb and axial gauges by
canonical gauge transformations and may worry about the equivalence of
the path integral, representing physical amplitude
in the holomorphic gauge to
those in the conventional ones. This worry can be taken care of by
the BRST formulation \cite{brst} as discussed in ref.\cite{lo1}: The
equivalence of the path integral in the holomorphic gauge to those in
conventional gauges follows from the Fradkin and Vilkovisky theorem
\cite{brst,fradkin}. In the frame-work of the coherent state quantization,
the holomorphic gauge condition is a legitimate gauge fixing condition.

Employing the coherent state quantization for the gauge fields,
we write the path integral representing the physical transition
amplitude \cite{fadsla} as
\[ Z=\int D p^z D q^{\bar z} D p^{\bar z} D
q^z D\cos\theta D\phi  D A_z D A_{\bar z} \delta(\Phi) \delta(\chi) \det
\{\chi,\Phi\}  \]
\begin{equation}
\exp\left\{
-\kappa i \int d^2z\left(A^f_{\bar z} A^f_z+A^i_{\bar z} A^i_z\right)\right\}
\exp\{i\int^{t_f}_{t_i} dt (K - H)\},\label{action}
\end{equation}
\[ K = \sum_\alpha\left(p^{\bar z}_\alpha \dot z_{\alpha}+
p^{z}_\alpha \dot{{\bar z}}_{\alpha}+
J_\alpha \cos \theta_\alpha \dot \phi_\alpha\right)
+\int d^2 z {\kappa\over 2}\left(\dot A^a_z A^a_{\bar z} -\dot A^a_{\bar z}
A^a_{ z}\right)
\]
where $\chi = 0$ is the gauge condition.
In the holomorphic gauge $\chi = A^a_{\bar z}$ and $\det \{\chi,\Phi\}= \det
\partial_{\bar z}$. Unless we consider a nontrivial topology for the two
dimensional space, $\det\partial_{\bar z}$ is not singular. Since it is
independent of dynamical variables, it can be dropped from the path integral.
Therefore we find that the path integral is now reduced to
\begin{equation}
Z  =\int D p^z D p^{\bar z} D
q^z D q^{\bar z} D\cos \theta D \phi
D A_z D A_{\bar z} \delta(A^a_{\bar z}) \delta
(\Phi^a) \exp\{i\int dt(K - H)\}.
\end{equation}

\section{Chern-Simons Quantum Mechanics in Holomorphic Gauge}
\label{csqm}

The quantum mechanical description of the NACS particles is obtained if we
integrate out the field variables. By use of the
given solution in the holomorphic gauge Eq.(\ref{sol}), we get the path
integral expressed only in terms of the quantum mechanical variables
\begin{equation}
Z  =\int D p^z D p^{\bar z} D
q^z D q^{\bar z} D\cos \theta D \phi
\exp\{i\int dt(K - H)\}, \label{qmpath}
\end{equation}
\[ K = \sum_\alpha\left(p^{\bar z}_\alpha \dot z_{\alpha}+
p^{z}_\alpha \dot{{\bar z}}_{\alpha}+
J_\alpha \cos \theta_\alpha \dot \phi_\alpha\right), \]
\[ H = \sum_\alpha {2\over m_\alpha}
 p^z_\alpha \left( p^{\bar z}_\alpha -   A^a_z (
z_\alpha,\bar z_\alpha)  Q^a_\alpha\right)  \]
where $A^a_z(z,\bar z)$ denotes the holomorphic gauge solution Eq.(\ref{sol}).

The operator formulation of the Chern-Simons quantum mechanics follows from
the observation that the path integral Eq.(\ref{qmpath}) is equivalently
expressed as
\begin{equation}
Z = <\eta_f\vert \exp\{ -i \hat{H}(t_f-t_i)\} \vert \eta_i>,\label{qmop}
\end{equation}
\[ \hat{H} = \sum_\alpha {2\over m_\alpha}
\hat p^z_\alpha \left(\hat p^{\bar z}_\alpha -  \hat A^a_z (
z_\alpha) \hat Q^a_\alpha\right).  \]
The variables
with  ` $\hat{}$ ' denote quantum operators of which commutators are given by
\[ [ {\bar z}_\alpha, \hat
p^{z}_\alpha] = i,\quad [{z}_\alpha, \hat p^{\bar z}_\alpha]= i,\quad
[\hat Q^a_\alpha,\hat Q^b_\beta] =i\epsilon^{abc} \hat
Q^c_\alpha \delta_{\alpha\beta}.\]
The gauge field $ \hat A^a_z$ here stands for the operator version of the
solution Eq.(\ref{sol}). Since the iso-vector operators $\hat Q^a$'s
satisfy the $SU(2)$ algebra,
they can be represented by $SU(2)$ some generators,
say, $T^a_j$ in a  representation of isospin $j$. For example, when $j=1/2$,
$\hat Q^a$'s are represented by the Pauli matrices $\tau^a/2$ and the state
vector $|\eta>$ by an iso-spin doublet. We suppressed the isospin indices in
Eq.(\ref{qmop}). In passing, note that $\hat Q^2_\alpha |\eta>=
J^2_\alpha|\eta> = j_\alpha(j_\alpha+1)|\eta>$, $j_\alpha\in {\bf Z}_{n+1/2}$.

We now conclude that the dynamics of the NACS particles are determined by
the Hamiltonian $\hat H$
\[ \hat {H} = -\sum_\alpha {1\over m_\alpha}\left(\nabla_{\bar
z_\alpha}\nabla_{z_\alpha}  +\nabla_{z_\alpha}\nabla_{\bar
z_\alpha}\right) \]
\begin{equation}
\nabla_{z_\alpha} ={\partial\over \partial z_\alpha}  +{1\over 2\pi
\kappa}\left( \sum_{\beta\not=\alpha}
\hat Q^a_\alpha \hat Q^a_\beta {1\over
z_\alpha -z_\beta}+\hat Q^2_\alpha a_z (z_\alpha)\right)\label{ham}
\end{equation}
\[\nabla_{\bar z_\alpha} ={\partial\over \partial \bar z_\alpha}\]
where $a_z (z_\alpha)=\lim_{z\rightarrow z_\alpha} 1/(z-z_\alpha)$ and it
needs an appropriate regularization.
The second term and the third term in $\nabla_{z_\alpha}$ describe mutual and
self interactions of NACS particles respectively. The Hamiltonian
Eq.(\ref{ham}) without the self interaction terms has been conjectured
and applied to the non-Abelian Aharonov-Bohm effect by Verlinde \cite{ver}.

The mutual interaction terms in $\hat H$ are  responsible for the non-Abelian
statistics. This becomes clear as we remove the interaction terms
by some singular non-unitary transformation, {\it i.e.} casting the wave
function $\Psi_h$ for the NACS particles in the holomorphic gauge into the
following form
\begin{equation}
\Psi_h(z_1,\dots,z_N) = U^{-1}(z_1,\dots,z_N)
\Psi_a (z_1,\dots,z_N),\label{wave}
\end{equation}
where $U^{-1}(z_1,\dots,z_N)$ satisfies the Knizhnik-Zamolodchikov (KZ)
equation \cite{kz}
\begin{equation}
\left({\partial\over \partial z_\alpha}  + {1\over 2\pi
\kappa} \sum_{\beta\not=\alpha} \hat Q^a_\alpha \hat Q^a_\beta {1\over
z_\alpha -z_\beta}\right) U^{-1}(z_1,\dots,z_N) =0.\label{kzeq}
\end{equation}
The Hamiltonian for $\Psi_a$ is a free Hamiltonian  $\hat H=
-\sum_\alpha \frac{ 2}{m_\alpha}
(\partial_{\bar z_\alpha}\partial_{z_\alpha})$.

In Eq.(\ref{wave}) we suppress the effects of the self interaction for the
following reason. The self interaction terms can be also removed in a similar
way as the mutual interaction terms are removed. As a result, we may have a
factor
\begin{equation}
\exp\left(-{1\over 2\pi \kappa}\sum_\alpha \lim_{z\rightarrow z_\alpha}\int^z
{\hat Q^2_\alpha \over z^\prime-z_\alpha} dz^\prime\right)\label{factor}
\end{equation}
in the r.h.s. of Eq.(\ref{wave}). This factor is not well defined until we
specify how to take the limit $z\rightarrow z_\alpha$. Here we take the limit
near $z_\alpha$ along $z=z_\alpha+t\epsilon_\alpha$ where $\epsilon_\alpha$
is a complex number with $|\epsilon_\alpha|=1$ and $t$ is a
real parameter to be taken $0$ in the limit.
Then the factor Eq.(\ref{factor}) is factorized into a divergent piece
\begin{equation}
\lim_{t\rightarrow 0}\prod_\alpha \exp\left(
-\frac{1}{2\pi\kappa}\hat Q^2_\alpha
\ln t\right)\label{facd}
\end{equation}
and a regular one
\begin{equation}
\prod_\alpha \exp\left(-\frac{i}{2\pi\kappa}\hat Q^2_\alpha
\arg\epsilon_\alpha\right).\label{facr}
\end{equation}
Since the divergent piece Eq.(\ref{facd}) can be absorbed into
a normalization
constant of the wave function,
the quantum mechanical Hamiltonian for the NACS
particles does not contain any singularity.
We may further absorb the constant
regular factor Eq.(\ref{facr}) into the phase of the wave function.
Therefore
we may safely remove the self interaction terms in the nonrelativistic
Hamiltonian Eq.(\ref{ham}) for the NACS
particles as in the case for the anyons
\cite{jackiw}. However, in the context of the relativistic quantum mechanics
for the NACS particles,
the regular factor Eq.(\ref{facr}) cannot be no longer
ignored: Noting that the regularization
procedure described above corresponds
to the framing of the knots \cite{witt} which represent
the world lines of the
relativistic NACS particles,
we may identify the anomalous spins of the NACS
particles as $2j_\alpha (j_\alpha+1)/4\pi\kappa$ from
the additional phase which
the wave function acquires due to the regular factor under $2\pi$ rotation.

The KZ equation has a formal solution which is
expressed as a path ordered line integral
in the $N$-dimensional complex space
\begin{equation}
U^{-1}(z_1,\dots,z_N) = P \exp\left[-{1\over 2\pi\kappa} \int_\Gamma
\sum_\alpha d\zeta^\alpha  \sum_{\beta\not=\alpha}
\hat Q^a_\alpha \hat Q^a_\beta
{1\over \zeta_\alpha -\zeta_\beta}\right],\label{kzsol}
\end{equation}
where $\Gamma$ is a path in the
$N$-dimensional complex space with one end point fixed (reference point) and
the other being $\zeta_f = (z_1,\dots,z_N)$. Explicit evaluation  of
the above formal expression will give the monodromy matrices.
We see that $\Psi_a$ obeys the braid statistics, due to the transformation
function $U(z_1,\dots,z_N)$ while $\Psi_h$ satisfies ordinary
statistics. In analogy with the Abelian Chern-Simons particle theory
we may call $\Psi_a$ the NACS particle wave function in the anyon gauge.
The two descriptions for the NACS particle -
one in the holomorphic gauge with
$\Psi_h$ and $H$ in Eq.(\ref{ham}) and the other in the anyon gauge with
$\Psi_a$ and the free Hamiltonian - are equivalent and the transformation
function between two gauges is given by  $U(z_1,\dots,z_N)$ Eq.(\ref{kzsol}).
It also defines the inner product in the holomorphic gauge
\begin{equation}
<\Psi_1 |\Psi_2> = \int d^{2N}\zeta \Psi_1(\zeta)^\dagger U^\dagger
(\zeta) U(\zeta) \Psi_2 (\zeta)
\end{equation}
where $\zeta = (z_1,\dots,z_N)$. The Hamiltonian in the holomorphic gauge
Eq.(\ref{ham}) is certainly Hermitian with this inner product
\[\nabla^\dagger_{z_\alpha}=-\nabla_{\bar z_\alpha},\quad
\nabla^\dagger_{\bar z_\alpha}=-\nabla_{z_\alpha},\quad
\hat H^\dagger = \hat H.\]

When $N=2$, the KZ equation is particularly simple and
the solution for the KZ
equation can be easily obtained
\begin{equation}
U(z_1,z_2)= \exp\left[\hat Q^a_1
\hat Q^a_2\frac{1}{2\pi\kappa}\ln(z_1-z_2)\right].
\end{equation}
Therefore if we exchange the positions
of two NACS particles along an oriented
path as depicted by Fig. 1, the wave function
in the anyon gauge transforms as
\begin{equation}
\Psi_a(z_1,z_2) \rightarrow \Psi_a(z_2,z_1) = \exp\left(\frac{\hat Q^a_1
\hat Q^a_2}{2\kappa}i\right) \Psi_a(z_1,z_2).
\end{equation}
The operator ${\cal R}_{\alpha\beta}=\exp\left(\frac{\hat Q^a_\alpha\hat
Q^a_\beta}{2\kappa}i\right)$ is called the braid operator or half-monodromy
which satisfies the Yang-Baxter equation and exhibits some character of the
non-Abelian statistics. If we wind one NACS particle around the other, {\it
i.e.} doubly exchange, the transformation of the wave function of the system
is given by the monodromy operator ${\cal M}_{\alpha\beta}=
({\cal R}_{\alpha\beta})^2$, \[\Psi_a(z_1,z_2) \rightarrow
\exp\left(\frac{\hat
Q^a_1 \hat Q^a_2}{\kappa}i\right) \Psi_a(z_1,z_2).\]

\section{Two Body Problem in Chern-Simons Quantum Mechanics}

Being equipped with the non-Abelian Chern-Simons quantum mechanics discussed
in the previous section, we now specifically consider a system of two NACS
particles. Although two body system does not reveal all the novel features of
the NACS particles, it certainly enables us to capture some essence of them.
Since two body problem is presently the only one known to have
explicit solutions, it is worth while to explore the two body system in
some detail. This section also serves as a preparatory stage for the
discussions on the non-Abelian Aharonov-Bohm effect in the following section.

The system of two NACS particles is described by the Hamiltonian
\[ \hat {H} = -\sum_{\alpha=1}^{2}{1\over m_\alpha}\left(\nabla_{\bar
z_\alpha}\nabla_{z_\alpha}  +\nabla_{z_\alpha}\nabla_{\bar
z_\alpha}\right) \]
\begin{equation}
\nabla_{z_1} ={\partial\over \partial z_1}  +\Omega{1\over
z_1 -z_2}, \quad
\nabla_{\bar z_1} ={\partial\over \partial \bar z_1},\label{two}
\end{equation}
\[ \nabla_{z_2} ={\partial\over \partial z_2}  +\Omega{1\over
z_2 -z_1}, \quad
\nabla_{\bar z_2} ={\partial\over \partial \bar z_2}. \]
Note that $\Omega$ is a block-diagonal matrix
\begin{equation}
\Omega= \hat Q^a_1\hat Q^a_2 / (2\pi\kappa)=\frac{1}{4\pi\kappa}
\left((\hat Q_1+\hat Q_2)^2-(\hat Q_1)^2-(\hat Q_2)^2\right)\label{omega}
\end{equation}
\[=\sum_j\frac{1}{4\pi\kappa}\left(j(j+1)-j_1(j_1+1)-j_2(j_2+1)\right)\otimes
{I}_j\]
and the half-monodromy is written as
\begin{equation}
{\cal R}=\exp(i\Omega\pi)\label{half}
\end{equation}
where $|j_1-j_2|\le j\le j_1+j_2$ and ${ I}_j$ is an identity matrix in
the subspace of total isospin $j$ which is spanned by $\{|m|\le j;
|j,m>\}$. In Eq. (\ref{omega}), $j(j+1)$, $j_1(j_1+1)$, and $j_2(j_2+1)$ are
eigenvalues of $(\hat Q_1+\hat Q_2)^2$, $(\hat Q_1)^2$, and $(\hat Q_2)^2$
respectively.

Upon introducing the center of mass and relative coordinates
$Z = (z_1+z_2)/2,\quad z = z_1 -z_2,$
the two body Hamiltonian Eq.(\ref{two}) becomes
\begin{equation}
\hat H = -{1\over 2\mu} \partial_Z \partial_{\bar Z}
-\frac{1}{\mu}(\nabla_z\nabla_{\bar z} +\nabla_{\bar z}\nabla_z),
\end{equation}
\[ \nabla_z = \partial_z +\frac{\Omega}{z},\quad \nabla_{\bar z} =
\partial_{\bar z}\]
where we take the mass of NACS particles, $m_1 = m_2 =2\mu$ for simplicity.
As usual, the motion of center of mass coordinates is not dynamical and
is decoupled from the motion of relative coordinate. If we write the wave
function for the two body system as $\Psi(z,\bar z, Z,\bar Z, t) = \Psi_{\rm
CM} (Z,\bar Z,t) \psi(z,\bar z,t)$, the Schr\"odinger equation for the system
becomes
\begin{equation}
-{1\over 2\mu} \partial_Z \partial_{\bar Z} \Psi_{\rm CM} (Z,\bar Z,t) =
i\frac{\partial \Psi_{\rm CM} (Z,\bar Z, t)}{\partial t}
\end{equation}
\begin{equation}
-\frac{1}{\mu}(\nabla_z\nabla_{\bar z} +\nabla_{\bar z}\nabla_z)
\psi(z,\bar z,t) = i\frac{\partial \psi(z,\bar z,t)}{\partial
t}.\label{hamrel}
\end{equation}

Since the center-of-mass Hamiltonian $\hat H_{\rm CM} =
-{1\over 2\mu} \partial_Z \partial_{\bar Z}$ has a set of eigenstates
$\{e^{i\left({\bf K}\cdot {\bf R}-(K^2/2\mu)t\right)}\}$,
the two body problem
now reduces to finding energy eigenfunctions $ \psi_n(z,\bar z)$ of the
Hamiltonian for the relative motion $\hat H_{\rm rel}$
\begin{equation}
\hat H_{\rm rel} = -\frac{1}{\mu}(\nabla_z\nabla_{\bar z} +
\nabla_{\bar z}\nabla_z),\label{eigen}
\end{equation}
\[\hat H_{\rm rel} \psi_n(z,\bar z) =
E_n  \psi_n(z,\bar z).\]
If the energy eigenfunctions $\psi_n(z,\bar z)$ are found, the energy
eigenstates of the system of two NACS particles are written as
\[\{\Psi_{K,n}(z,\bar z, Z,\bar Z, t) =
e^{-i\left(E_n+K^2/2\mu\right)t} e^{i{\bf K}\cdot
{\bf R}}\psi_n(z,\bar z)\}.\]

To obtain the energy eigenfunctions for $\hat H_{\rm rel}$ we find it useful
to take a similarity transformation given by
\begin{equation}
\hat H_{\rm rel} \longrightarrow  \hat H_{\rm rel}^\prime =
G^{-1} \hat H_{\rm rel} G,
\end{equation}
\[\psi_n(z,\bar z) \longrightarrow
\psi_n^\prime(z,\bar z) =G^{-1} \psi_n(z,\bar z).\]
The following observations help us to solve the Schr\"odinger equation for
the relative motion. Firstly, the transformed Hamiltonian $\hat H_{\rm
rel}^\prime$ has the same set of eigenvalues $\{E_n\}$ as $\hat H_{\rm rel}$
\begin{equation}
\hat H_{\rm rel}^\prime \psi_n^\prime(z,\bar z) = E_n
\psi_n^\prime(z,\bar z).\label{treigen}
\end{equation}
Secondly, if we choose the transformation function $G$ to be
\begin{equation}
G (z,\bar z) = \exp\left(-\frac{\Omega}{2}\ln(z\bar z)\right),
\end{equation}
the inner product between the states becomes usual and
transformed Hamiltonian $\hat H_{\rm rel}^\prime$
becomes manifestly Hermitian
\begin{equation}
\hat H_{\rm rel}^\prime = -\frac{1}{\mu}
(\nabla_z^\prime\nabla_{\bar z}^\prime +
\nabla_{\bar z}^\prime\nabla_z^\prime),\label{trham}
\end{equation}
\[ \nabla_z^\prime=\partial_z + \frac{\Omega}{2}\frac{1}{z},\quad
\nabla_{\bar z}^\prime=\partial_{\bar z}
-\frac{\Omega}{2}\frac{1}{\bar z}.\]

The Hamiltonian $\hat H_{\rm rel}^\prime$ resembles the Hamiltonian for the
relative motion of two anyon system in Coulomb gauge. Indeed it may be
regarded as the Hamiltonian for the system of two NACS particles in Coulomb
gauge in a sense. As we mentioned earlier, in general, the Gauss' constraint
cannot be solved explicitly in Coulomb gauge
in that the constraint is nonlinear
in this gauge. However, this is not the case for the one particle sector.
Suppose that one NACS particle is located at the origin. One can easily find
that the Gauss' constraint Eq.(\ref{gaus}) which is now simplified to be
\begin{equation}
\Phi^a = {\kappa \over 2}\epsilon^{ij} F^a_{ij} ({\bf x}) +
Q^a \delta({\bf x}) =0\label{gausone}
\end{equation}
has an explicit solution in the Coulomb gauge
\begin{equation}
A^a_i ({\bf x}) = \epsilon_{ij} \frac{Q^a}{2\pi\kappa}\frac{x^j}{r^2},\quad
\partial_i A^a_i ({\bf x}) =0.
\end{equation}
Therefore, if we neglect the self interaction, we can construct a Hamiltonian
for the system of two NACS particles as follows
\begin{equation}
\hat H_C=\sum_{\alpha=1, 2} {1\over 2 m_\alpha}\left(p^i_\alpha-A^{ai}({\bf
q}_\alpha) Q^a_\alpha\right)^2,
\end{equation}
\[
A^a_i ({\bf q}_1) =\epsilon_{ij}\frac{Q^a_2}{2\pi\kappa}\frac{q^j_1-q^j_2}
{\left({\bf q}_1-{\bf q}_2\right)^2},\]
\[
A^a_i ({\bf q}_2) =-\epsilon_{ij}\frac{Q^a_1}{2\pi\kappa}\frac{q^j_1-q^j_2}
{\left({\bf q}_1-{\bf q}_2\right)^2}.\]
Certainly, the Hamiltonian for the relativistic motion which
follows from $\hat
H_C$ coincides with $\hat H^\prime_{\rm rel}$.
This shows that the description
of the two-body system in the test particle frame-work is valid.

Rewriting the Hamiltonian $\hat H_{\rm rel}^\prime$ in the polar coordinates
and projecting it onto the subspace of total isospin $j$, we have
\begin{equation}
\hat H_{\rm rel}^\prime =
-\frac{1}{2\mu}\left[\frac{\partial^2}{\partial r^2}+
\frac{1}{r}\frac{\partial}{\partial r}+\frac{1}{r^2}\left(\frac{\partial}
{\partial \theta}+i\omega_j\right)^2\right].\label{hampr}
\end{equation}
where $\Omega|j,m>=\omega_j|j,m>$ and
\begin{equation}
\omega_j=\frac{1}{4\pi\kappa}
\left(j(j+1)-j_1(j_1+1)-j_2(j_2+1)\right).\label{eigenv}
\end{equation}
As is well-known in the case of Aharonov-Bohm scattering \cite{ahbo}, the
Hamiltonian $\hat H_{\rm rel}^\prime$ has eigenfunctions
\[\{ n\in {\bf
Z}; e^{in\theta} J_{n+\omega_j}(kr),  e^{in\theta} J_{-n-\omega_j}(kr)\}\]
with an eigenvalue $E_k = \frac{\hbar^2 k^2}{2\mu}$.
Requiring the solution to
be regular at the origin, we get a set of eigenfunctions
\begin{equation}
\{ n\in {\bf Z}; e^{in\theta} J_{|n+\omega_j|}(kr)\}.\label{col}
\end{equation}
Subsequently, we have eigenfunctions in the holomorphic gauge
\begin{equation}
\{ n\in {\bf Z}; \exp\left(-\frac{\omega_j}{2}\ln z\bar
z\right) e^{in\theta} J_{|n+\omega_j|}(kr)\}\label{hol}
\end{equation} and eigenfunctions in the anyon gauge
\begin{equation}
\{ n\in {\bf Z}; \exp\left(i\omega_j\theta\right) e^{in\theta}
J_{|n+\omega_j|}(kr)\}.\label{any}
\end{equation}
Note that the eigenfunctions in the anyon gauge are multivalued.

\section{Non-Abelian Aharonov-Bohm Effect}
\label{sec:naab}

Having obtained the Hamiltonian for the relative motion Eq.(\ref{col}),
we can
proceed to our discussion on the non-Abelian Aharonov-Bohm scattering in
parallel to the discussion on the Aharonov-Bohm scattering in ref.
\cite{ahbo}.
 From the discussion on the two body system in
the previous section, it is clear that the state of the system can be
characterized by $|{\bf k}>\otimes |j_1,j_2;m_1,m_2>$ where  $|{\bf k}>$ and
$|j_1,j_2;m_1,m_2>$ describe the momentum and isospin state of the system
respectively.

Now let us examine the structure of the Hamiltonian
$\hat H_{\rm rel}^\prime$, Eq.(\ref{hampr}). The Hamiltonian commutes with
the operators $(\hat Q_1+\hat Q_2)^2$, $(\hat Q_1)^2$, and $(\hat Q_2)^2$ and
it implies that the scattering amplitude $f({\bf k}^\prime, {\bf k})$ has the
following structure
\begin{equation}
f({\bf k}^\prime, {\bf
k})_{\{m_1^\prime,m_2^\prime;m_1,m_2\}}=\sum_{j=|j_1-j_2|}^{j_1+j_2}
\sum_{m=-j}^{j}\label{scatt}
\end{equation}
\[<j_1,j_2;m_1^\prime,m_2^\prime|j_1,j_2;j,m>f_j
({\bf k}^\prime, {\bf k}) <j_1,j_2;j,m|j_1,j_2;m_1,m_2> \]
where
$<j_1,j_2;j,m|j_1,j_2;m_1,m_2>$ is the well-known Clebsch-Gordan coefficient.
We may call $f_j ({\bf k}^\prime, {\bf k})$ the scattering amplitude of
isospin $j$ channel.
Here we assume that two NACS particles are distinguishable and will consider
the collision between indistinguishable NACS particles after a little.

In order to evaluate $f_j ({\bf k}^\prime, {\bf k})$,
we only need to consider
the channel of isospin $j$ where the dynamics of the system is governed by
$\hat H_{\rm rel}^\prime$, Eq.(\ref{hampr}). The most general solution of the
Schr\"odinger equation
\begin{equation}
\hat H_{\rm rel}^\prime \psi = \frac{\hbar^2 k^2}{2\mu} \psi
\end{equation}
is given as a superposition of the eigenfunctions Eq.(\ref{col})
\begin{equation}
\psi = \sum_n a_n e^{in\theta} J_{|n+\omega_j|}(kr) \label{gen}
\end{equation}
and the scattering solution corresponds to the solution which has a
spatial asymptotics of
\begin{equation}
e^{ikr\cos\theta-i\omega_j \theta} +
f_j(k,\theta)\frac{e^{ikr}}{\sqrt{r}}.\label{scat}
\end{equation}
Fig. 2 depicts scattering of two NACS particles in the center of mass frame.
Due to the long range nature of the Aharonov-Bohm effect, the portion
of the incident wave is not a plane wave as in the case of the Coulomb
scattering \cite{gott}.

Comparing the asymptotics of the solution Eq.(\ref{gen})
\begin{equation}
\psi \longrightarrow \sum_n a_n e^{in\theta} \sqrt{\frac{2}{\pi kr}}
\cos\left(kr-\frac{|n+\omega_j|\pi}{2}-\frac{\pi}{4}\right)
\end{equation}
with the asymptotics of the expansion of the plane wave in angular momentum
eigenstates
\begin{equation}
e^{ikr\cos\theta}=\sum_n i^n e^{in\theta} J_n(kr)
\end{equation}
\[\longrightarrow\sum_n i^n e^{in\theta} \sqrt{\frac{2}{\pi kr}}
\cos\left(kr-\frac{n\pi}{2}-\frac{\pi}{4}\right)\]
suggests that $a_n = \exp\left[\pi i(n- |n+\omega_j|/2)\right]$.
Thus the scattering solution is
\begin{equation}
\psi=\sum^\infty_{n=-[\omega_j]} e^{i\pi(n-\omega_j)/2} e^{in\theta}
J_{n+\omega_j}(kr)+
\end{equation}
\[\sum^{-[\omega_j]-1}_{-\infty }e^{-i\pi(n-\omega_j)/2}
e^{in\theta}J_{-n-\omega_j}(kr).\]

We can obtain the spatial asymptotics of $\psi$ either by use of a
differential equation \cite{ahbo,hagen} or by use of the Schl\"afli contour
representation of Bessel function \cite{jackiw}
\begin{equation}
\psi \longrightarrow e^{ikr\cos\theta-i\omega_j\theta}
+\frac{e^{-i\omega_j\pi}}{\sqrt{2\pi k i}} e^{-i\left([\omega_j]+\frac{1}{2}
\right)\theta}\frac{\sin\omega_j\pi}{\sin\theta
/2}\frac{e^{ikr}}{\sqrt{r}}.\label{asym}
\end{equation}
Then the scattering amplitude of isospin $j$ channel \cite{comm} is read as
follows \begin{equation}
f_j(k,\theta)= \frac{e^{-i\omega_j\pi}}{\sqrt{2\pi k i}}
e^{-i\left([\omega_j]+\frac{1}{2}\right)\theta}\frac{\sin\omega_j\pi}
{\sin\theta/2}.\label{scatj}
\end{equation}

The scattering amplitude $f_j ({\bf k}^\prime, {\bf k})$ follows from
Eq.(\ref{scatt}) and Eq.(\ref{scatj}). This result may be expressed more
succinctly if we define a matrix ${\cal F}({\bf k}^\prime, {\bf k})$ in the
isospin space
\begin{equation}
f({\bf k}^\prime, {\bf
k})_{\{m_1^\prime,m_2^\prime;m_1,m_2\}}=
<j_1,j_2;m_1^\prime,m_2^\prime|{\cal F}({\bf k}^\prime, {\bf k})
|j_1,j_2;m_1,m_2>.
\end{equation}
(Since $j_1$ and $j_2$ are fixed, they will be omitted in the formulae
hereafter.)
We find that ${\cal F}({\bf k}^\prime, {\bf k})$ can be written as
\begin{equation}
{\cal F}({\bf k}^\prime, {\bf k})= \frac{e^{-i\Omega\pi}}{\sqrt{2\pi k i}}
e^{-i\left([\Omega]+\frac{1}{2}\right)\theta}
\frac{\sin\Omega\pi}{\sin\theta/2}\label{scatm}
\end{equation}
where $[\alpha]$ denotes the integer such that $0\le \alpha -[\alpha]< 1$
and $[\Omega]$ should be understood as $[\Omega]|j,m>=[\omega_j]|j,m>$.
In terms of the half-monodromy matrix Eq.(\ref{half}) or monodromy
${\cal M}={\cal R}^2$, it may be
expressed as
\begin{equation}
{\cal F}({\bf k}^\prime, {\bf k})=\frac{1}{\sqrt{2\pi k i}}{\cal R}^{-1}
({\cal R}^{-1}-{\cal R})
\frac{e^{-i[\Omega]\theta}}{1-e^{i\theta}}\label{scatmm}
\end{equation}
\[=\frac{1}{\sqrt{2\pi k i}}({\cal M}^{-1}-1)
\frac{e^{-i[\Omega]\theta}}{1-e^{i\theta}}.\]
This result is similar to that obtained by Lo and Preskill \cite{lopre}, but
does not completely agree with. Note in particular
that the obtained scattering
amplitude Eq.(\ref{scatm}) is single valued in contrast to theirs.
However, as we shall see, both scattering
amplitudes yield the same `inclusive' cross section.

The differential cross section for the scattering process
$|m_1,m_2>\rightarrow |m^\prime_1,m^\prime_2>$ is then given by
\begin{equation}
\frac{d\sigma}{d\theta}\left(|m_1,m_2>\rightarrow |m^\prime_1,m^\prime_2>
\right) = \Bigm|<m^\prime_1,m^\prime_2|{\cal F}(k,\theta)|m_1,m_2>\Bigm|^2.
\end{equation}
Writing down its explicit expression, we have
\begin{equation}
\frac{d\sigma}{d\theta}\left(|m_1,m_2>\rightarrow |m^\prime_1,m^\prime_2>
\right) =-\frac{1}{8\pi k} \sum_{j,m}\sum_{j^\prime, m^\prime}
<m^\prime_1,m^\prime_2|j,m>
\end{equation}
\[<j,m|m_1,m_2>
<m_1,m_2|j^\prime,m^\prime><j^\prime,m^\prime|m^\prime_1,m^\prime_2>\]
\[\left(e^{-2i\omega_j\pi}-1\right)\left(e^{-2i\omega_{j^\prime}\pi}-1\right)
\frac{e^{-i([\omega_j]-[\omega_{j^\prime}])\theta}}{\sin^2\theta/2}.\]
If we do not measure the isospin orientations of particles after scattering,
the cross section is given by
\begin{equation}
\frac{d\sigma}{d\theta}\left(|m_1,m_2>\rightarrow {\rm all}\right)
=<m_1,m_2|{\cal F}^\dagger(k,\theta){\cal F}(k,\theta)|m_1,m_2>\label{incl}
\end{equation}
\[=\frac{1}{2\pi k} \frac{<m_1,m_2|\sin^2 \Omega\pi|m_1,m_2>}{\sin^2
\theta/2}\]
\[=\frac{1}{4\pi k} \frac{1}{\sin^2 \theta/2}\left(1-{\rm
Re}<m_1,m_2|{\cal M} |m_1,m_2>\right).\]
This is called inclusive cross section in ref.\cite{lopre}. The formula for
the inclusive cross Eq.(\ref{incl}) was obtained first by Verlinde \cite{ver}
and was confirmed recently by Lo and Preskill \cite{lopre}.

If two NACS particles share the same physical properties and belong to
the same isospin multiplet, {\it i.e.} $j_1=j_2$ (but $m_1$ may differ
from $m_2$), it is appropriate to regard them indistinguishable
\cite{lopre}. In such a case, two body
system is described by the symmetrized wave function
\begin{equation}
\Psi(z,\bar z, Z,\bar Z, t) = \frac{1}{\sqrt{2}}\Psi_{\rm
CM} (Z,\bar Z,t)\left(\psi(z,\bar z,t) +\psi(-z,-\bar
z,t)\right).\label{symm}
\end{equation}
(We describe the NACS particles
in the holomorphic gauge as bosons interacting
with each other through the non-Abelian
Chern-Simons gauge field. Alternatively
we may describe the NACS particles in terms of the fermions, then we must
choose the antisymmetrized wave function.)
Going through the same analysis given in sections V and VI with the
symmetrized wave function Eq.(\ref{symm}), we find that the scattering
amplitude $f_j^I ({\bf k}^\prime, {\bf k})$
for the scattering of indistinguishable NACS particles in the
isopsin $j$ channel is given in terms of $f_j ({\bf k}^\prime, {\bf k})$
Eq.(\ref{scatj}) by
\begin{equation}
f_j^I (k,\theta)= f_j(k,\theta)+ f_j(k,\pi-\theta)
\end{equation}
\[ =\frac{e^{-i\omega_j\pi}}{\sqrt{2\pi k i}}
e^{-i\left([\omega_j]+\frac{1}{2}\right)\theta}\frac{\sin\omega_j\pi}
{\sin\theta/2}\left(1+e^{i\left([\omega_j]+1/2\right)(2\theta-\pi)}\tan
\theta/2\right).\]\label{scatb}
Then we obtain the scattering amplitude matrix
\begin{equation}
{\cal F}^I({\bf k}^\prime, {\bf k})=
{\cal F}({\bf k}^\prime, {\bf k})+{\cal F}(-{\bf k}^\prime, {\bf k})
\end{equation}
\[ =\frac{1}{\sqrt{2\pi k i}}({\cal M}^{-1}-1)
\frac{e^{-i[\Omega]\theta}}{1-e^{i\theta}}\left(1+
e^{i\left([\Omega]+1/2\right)(2\theta-\pi)}\tan
\theta/2\right)\]
\[ = {\cal F}({\bf k}^\prime, {\bf k}) \left(1+
e^{i\left([\Omega]+1/2\right)(2\theta-\pi)}\tan
\theta/2\right).\]
This yields the differential cross section for the indistinguishable NACS
particles
\begin{equation}
\frac{d\sigma^I}{d\theta}\left(|m_1,m_2>\rightarrow {\rm all}\right)=
<m_1,m_2|{\cal F}^{I\dagger}(k,\theta)
{\cal F}^I(k,\theta)|m_1,m_2>\label{incli}
\end{equation}
\[
=<m_1,m_2|{\cal F}^{\dagger}(k,\theta){\cal
F}(k,\theta)\left\{\sec^2\theta/2-2\cos\left((2[\Omega]+1)\theta\right)
\tan\theta/2\right\}|m_1,m_2>.
\]

\section{Conclusions and Discussion}
\label{concl}

Description of NACS particles often involves field degrees of
freedom, since they are usually described as isospin particles of
which isospin charges are coupled to non-Abelian Chern-Simons
gauge fields. However, it is not necessary to rely upon the non-Abelian
Chern-Simons gauge fields to discuss the quantum mechanics of NACS
particles. The non-Abelian Chern-Simons gauge fields are to be completely
determined in principle in terms of the quantum mechanical degrees of
freedom of the isopsin particles through the Gauss's law constraints, i.e.,
they themselves do not have dynamical degrees of freedom. Their only role is
to introduce a topological interaction between the isospin particles,
endowing them with non-Abelian magnetic fluxes. Thus, if we solve the Gauss'
law constraints explicitly, we are able to describe the dynamics of NACS
particles solely by a quantum mechanical Hamiltonian.

In this paper we have shown that the Gauss' law constraints can be solved
explicitly in the framework of coherent state quantizaion
in an appropriately chosen gauge and discussed the
dynamics of the NACS particles, based upon the obtained quantum mechanical
Hamiltonian, especially the non-Abelian Aharonov-Bohm
scattering: In the framework of the coherent state quantization, we are able
to choose the holomorphic gauge condition to fix the gauge degrees of
freedom thanks to the enlarged gauge orbit space and obtain the quantum
mechanical Hamiltonian which governs the dynamics of the NACS particles
through the KZ equation entirely in the holomorphic sector.
The obtained quantum mechanical Hamiltonian enabled us to discuss the
scattering of NACS particles in parallel to the discussion of the
Aharonov-Bohm scattering. Evaluating the differential scattering cross
section, we confirmed, yet in a more concrete way, the results which were
previously obtained by Verlinde \cite{ver} and by Lo and Preskill
\cite{lopre}. However, scattering amplitude which we obtained differs from
theirs in details, in particular, it is single valued in contrast to that of
Lo and Preskill. We conceive that the gauge condition chosen implicitly
in ref.\cite{lopre} be different from ours and it may make such a
difference. A close comparison between the two approaches would be
worthwhile.

Incorporating an additional $U(1)$ electromagnetic field into the NACS
particle theory, we can construct the model which has been proposed to
describe some fractional quantum Hall states called singlet quantum Hall
effect states \cite{wen,moo,balfra}. The non-Abelian quantum mechanics
developed in this paper will be certainly useful to explore their physical
properties. The proposed non-Abelian Chern-Simons quantum mechanics also
gives us immediate helps in discussing \cite{lo,tlee93} some subjects,
in more generalized context, which have been previously discussed in the
anyon theory such as construction of exact many-body wavefunctions
\cite{exact} and evaluation of the second virial coefficient \cite{virial}.
Although only the NACS particles with $SU(2)$ symmetry group is discussed in
this paper, our discussion on the NACS particles is not limited to the case
of $SU(2)$. Many of formal properties of NACS particles with $SU(2)$ are
expected to be shared among those with other symmetry groups. For an
example, the formal expression of the scattering matrix in terms of the
monodromy Eq.(\ref{scatmm}) may hold in general. A
construction of the quantum mechanical Hamiltonian for NACS particles with
$SU(N)$, $N\ge 2$ symmetry group is presented in ref.\cite{lo2} and
analyses on the NACS particles with more general symmetry groups are
currently in progress \cite{lo3}.
As we mentioned in the introduction, the NACS particles make their
appearances in various circumstances. For instance, they may appear as
gravitationally interacting point-like sources in (2+1) dimensions, as
non-Abelian vortices in (2+1) dimensions, or cosmic string with some
discrete non-Abelian charges in (3+1) dimensions. Since the coherent
state quantization and the holomorphic gauge fixing apply also to the
models describing those NACS particles of various kinds, it will be
certainly interesting to develop the non-Abelian Chern-Simons quantum
mechanics for them and to discuss the related subjects in the framework
given in this paper.

\acknowledgements

TL was supported in part by the KOSEF and PO was supported by the
KOSEF through  C.T.P. at S.N.U. TL would like to thank professor R.
Jackiw for valuable comments. We also thank professor C. Lee and
professor Y. M. Cho for useful discussions.

\figure{Exchange of two NACS particles along an oriented path.}
\figure{Scattering of two NACS particles in the center of mass frame.}

\end{document}